\documentclass[prd,aps,preprint,showpacs,nofootinbib,eqsecnum,superscriptaddress]{revtex4-1}
\usepackage{hyperref}
\usepackage{amsfonts,amssymb,amsmath,amsthm,amsxtra,mathtools,euscript,mathrsfs}

\usepackage{bm}%----------bold math text
\usepackage{graphicx,color,xcolor,subfigure}
\usepackage{dcolumn}%----------Align table columns on decimal point
\usepackage{bbm}

\def\a{\alpha}
\def\b{\beta}
\def\g{\gamma}
\def\d{\delta}
\def\e{\eta}
\def\l{\lambda}
\def\r{\rho}

\def\k{\kappa}

\def\m{\mu}
\def\n{\nu}
\def\f{\phi}

\def\o{\omega}
\def\O{\Omega}
\def\pa{\partial}
\def\na{\nabla}
\def\gmn{g_{\mu\nu}}
\def\gumn{g^{\mu\nu}}
\def\emn{\eta_{\mu\nu}}

\def\la{\mathcal{L}}

\def\exp{\mathrm{e}}

\def\mft{\mathfrak{T}}
\def\ti{\tilde}
\def\tgmn{\tilde{g}_{\m\n}}
\def\tgumn{\tilde{g}^{\m\n}}

\def\rb{\bar{r}}
\def\rt{\tilde{r}}

\newcommand{\etal}{\textit{et al}.}

\begin{document}
\title{Thermodynamics of $f(R)$ Gravity with Disformal Transformation}

%---author 1
\author{Chao-Qiang Geng}
\email[Electronic address: ]{geng@phys.nthu.edu.tw}
\affiliation{Synergetic Innovation Center for Quantum Effects and Applications (SICQEA),
 Hunan Normal University, Changsha 410081}
 \affiliation{Department of Physics,
National Tsing Hua University, Hsinchu 30013}
\affiliation{Physics Division,
National Center for Theoretical Sciences, Hsinchu 30013}
%---author 2
\author{Wei-Cheng Hsu}
\email[Electronic address: ]{etnn1234@gmail.com}
\affiliation{Department of Physics,
National Tsing Hua University, Hsinchu 30013}
%---author 3
\author{Jhih-Rong Lu}
\email[Electronic address: ]{jhih-ronglu@gapp.nthu.edu.tw}
\affiliation{Department of Physics,
National Tsing Hua University, Hsinchu 30013}
%---author 4
\author{Ling-Wei Luo}
\email[Electronic address: ]{lwluo@gate.sinica.edu.tw}
\affiliation{Department of Physics,
National Tsing Hua University, Hsinchu 30013}
\affiliation{Institute of Physics, Academia Sinica, Taipei 11529}

\begin{abstract}
We study thermodynamics in $f(R)$ gravity with the disformal transformation.
The transformation applied to the matter Lagrangian has the form of
$\g_{\m\n} = A(\phi,X)g_{\m\n} + B(\phi,X)\pa_\m\f\pa_\n\f$
with the assumption of the Minkowski matter metric $\g_{\m\n} = \e_{\m\n}$, 
where $\phi$ is the disformal scalar and $X$ is the corresponding kinetic term of $\phi$.
We verify the generalized first and second laws of thermodynamics 
in this disformal type of $f(R)$ gravity in the 
Friedmann-Lema\^{i}tre-Robertson-Walker (FLRW) universe.
In addition, we show that the Hubble parameter contains the disformally induced
terms, which define the effectively varying equations of state for matter.
\end{abstract}

\maketitle

%%%%%%%%%%%%%%%%%%%%%%%%%%%%%%%%%%%%%%%%%%%%%%%%%%%%%%%%%%%%%%%%%%%%%%%%%%%%%%%%%%%%%%%%%%%%%%%%%%%%%%%%%%%%%%%
%-----------------------------------------------------------------------%
% Sec. I - Introduction
%-----------------------------------------------------------------------%

\section{Introduction}

The connection between thermodynamics and general relativity (GR) 
has been found by studying black hole entropy.
In 1972, Bekenstein stated that this entropy is proportional to the 
area of the event horizon~\cite{Bekenstein:1973ur}.
The thermodynamical behavior of black holes was also examined in 1974 by
Bardeen, Carter, and Hawking in Ref.~\cite{Bardeen:1973gs}, 
showing that black hole entropy and temperature are associated with
the corresponding area, $A$, and surface gravity, $\kappa_s$, 
on the horizon, respectively.
In 1975, Hawking further presented that the 
proportionality of black hole temperature and surface gravity 
is equal to $1/{2\pi}$, i.e., $T=\kappa_s/{2\pi}$, by considering 
matter near the black hole horizon as quantum matter~\cite{Hawking:1974sw}.
%{\color{red}In 1990s, Wald studied black hole in generalized theory of gravity and proposed the dynamical entropy~\cite{Wald:1993nt,Iyer:1994ys}.}

%-----------------------------------------------------------------------%
%The application of this concept to cosmology was first done 
%by Gibbons and Hawking~\cite{Gibbons:1977mu}.
%They consider the de Sitter space with a positive cosmological constant~\cite{Einstein:1917ce}
%to ensure the existence of the event horizon of the universe.
%They verified that such approach satisfied the first and second laws of thermodynamics in cosmology.
%The derivation of the first law of thermodynamics by using the apparent horizon of the hypersurface of $(n + 1)$-dimensional FLRW universe has been shown in~\cite{Cai:2005ra}. 
%It is argued that for the dynamical universe, the apparent horizon is associated with gravitational entropy and surface gravity.

%-----------------------------------------------------------------------%
In 1995, Jacobson pointed out that the 
Einstein equation can be deduced from the thermodynamic properties
of spacetime together with the proportional relation
of the entropy and horizon area, 
which gives a deeper connection between thermodynamics and gravity~\cite{Jacobson:1995ab}.
Later, this idea was applied to cosmology. In particular, in 2005,
Cai and Kim~\cite{Cai:2005ra} demonstrated that the
Friedmann equations can be derived by applying thermodynamic
properties to the apparent horizon of the universe. 
Upon replacing different entropy formulae of the black hole
in different gravity theories, such as Gauss-Bonnet and Lovelock
gravity, one can obtain the corresponding modified Friedmann equations
~\cite{Cai:2005ra,Akbar:2006er}. 

%----------------------------------------------------------------------%

%{\color{red}
It is generally believed that studies of the connections between
	thermodynamics and gravity theories would give
	us some insight into the real nature of gravity.
	In particular, 
	gravity theories, such as  
	scalar-tensor~\cite{Akbar:2006er,Wu:2008ir,Cai:2006rs},
	$f(R)$~\cite{Akbar:2006er,Eling:2006aw,Akbar:2006mq,Wu:2008ir,Bamba:2009id,Bamba:2009ay,Bamba:2010kf}, 
	Gauss-Bonnet and Lovelock gravity~\cite{Cai:2006rs,Cai:2008mh,Paranjape:2006ca} and braneworld~\cite{Wu:2007se,Sheykhi:2007zp,Sheykhi:2007gi,Wu:2007em} models have been widely discussed for this purpose.
	It is known that $f(R)$ gravity, which takes the gravity part of
	the LaGrangian as a function of the Ricci scalar $R$ rather 
	than the LaGrangian given by Hilbert, is one of the popular modified gravity theories for understanding dark energy 
	in cosmology.
	% is called $f(R)$ gravity, which takes the gravity part of
	%the LaGrangian as a function of the Ricci scalar $R$ rather 
	%than the LaGrangian given by Hilbert.
	The field equation for $f(R)$ has also been derived 
	by considering spacetime as a non-equilibrium thermodynamic system
	such that an entropy production term is added
	in the Clausius relation, i.e., $d\hat{S}=dQ/T +d_i \hat{S}$.~\cite{Eling:2006aw}. Here, the horizon entropy is
	defined by $\hat{S}= F(R)A/(4G)$ with $F(R)=\partial f/
	\partial R$ and $d_i \hat{S}$  a bulk viscosity
	entropy production term.
	However, in the Friedmann-Lema\^{i}tre-Robertson-Walker (FLRW) universe, 
	the Friedmann equation  can be viewed both in 
	equilibrium and non-equilibrium thermodynamic descriptions ~\cite{Bamba:2009id}.
	%}

%-----------------------------------------------------------------------%
%{\color{red}
In addition to thermodynamical properties,
$f(R)$ gravity has been investigated in a wide variety of aspects.
	For example, the effect deviated from GR can be identified 
	as the effective dark energy,
	which could lead to the accelerating universe~\cite{Riess:1998cb,Perlmutter:1998np}.
	In addition, by choosing $f(R)=R^N$, one can show that 
	there is a correspondence between the 
	Einstein-conformally invariant Maxwell solutions and
	the solutions of $f(R)$ gravity without matter field~\cite{Hendi:2009sw}.
	Considering the trace anomaly as the source in $f(R)$ gravity, it
	can be demonstrated that there exist different (Schwarzschild-AdS (dS),
	Schwarzschild, Reissner-Nordström) black hole solutions 
	in different models~\cite{Hendi:2012nj}.
	Furthermore, $f(R)$ gravity has instanton solutions in 
	4-dimension Eguchi-Hanson space, and soliton solutions
	in 5-dimension Eguchi-Hanson-like spacetimes~\cite{Hendi:2012zg}.
	%}

%-----------------------------------------------------------------------%
%{\color{red}
After performing a conformal transformation on $f(R)$
gravity or a scalar-tensor theory, the two theories 
both become GR with a dynamical scalar field. 
In this sense, the two frames are mathematically equivalent.
However, the intriguing question is whether these
two frames are physically equivalent or not.
Capozziello \etal~ have showed that the two frames are physically
non-equivalent by considering specific $f(R)$ models in cosmology
and found that their Hubble parameters are different~\cite{Capozziello:2010sc}.
Similar situation happens when considering the finite time
cosmological singularities of $f(R)$ gravity. The
singularities change from one type to another when
transforming from one frame to the other~\cite{Bahamonde:2016wmz}.
The equivalence of both frames for the scalar-tensor theory 
have also been studied from the thermodynamics viewpoint~\cite{Bhattacharya:2017pqc}.
%}

%-----------------------------------------------------------------------%
%{\color{red}
Another alternative of modified gravitational theories is to
modify  Riemannian geometry into Finslerian one~\cite{Lammerzahl:2018lhw,Lammerzahl:2012kw,Itin:2014uia,Girelli:2006fw}.
Within this framework, physical equations such as Maxwell's equations,
and Dirac equations should be rewritten into a flat Finslerian
spacetime rather than a Minkowskian one~\cite{Lammerzahl:2018lhw,Itin:2014uia}.
In ~\cite{Bekenstein:1992pj}, Bekenstein considered a kind of
gravitational theories, which contains two geometries.
He allowed the physical geometry ($\gamma_{\mu\nu}$) with the matter dynamics
 could be Finslerian. To respect the
weak equivalence principle and causality, Finsler geometry must go back to Riemannian one, and the corresponding
matter metric, $\gamma_{\mu\nu}$, must be related to the gravitational metric
($g_{\mu\nu}$) by the so-called disformal transformation
\begin{equation}%%%%%%%%%%%
\g_{\m\n} = A(\phi, X)g_{\m\n} + B(\phi, X)\pa_\m\f\pa_\n\f \label{disformal Def.}\,,
\end{equation}
where $\f$  and $X$ are the disformal field and corresponding 
kinetic term, while $A$ and $B$ are functions of $\phi$ and $X$, 
respectively.
%}
This kind of the theory applied to relativistic
cosmology for the early universe was first done by Kaloper~\cite{Kaloper:2003yf},
in which the disformal field is considered to be the source of inflation.
The transformation can only couple to matter.
%{\color{red}
Motivated by the work in Ref.~\cite{Bekenstein:1992pj}, we consider
the situation that the flat Finslerian spacetime is reduced to the 
Minkowskian one, and examine thermodynamic properties in $f(R)$ gravity.
That is, in our study, we explore  thermodynamic 
properties in $f(R)$ gravity by including the disformal transformation with the assumption of the 
Minkowski matter metric, $\g_{\m\n}=\eta_{\m\n}$.
%}

%-----------------------------------------------------------------------%
The paper is organized as follows. In Sec.~\uppercase\expandafter{\romannumeral 2}, 
we first calculate the equations of motion of our model 
in the Jordan frame. Then, we derive the generalized first 
and second laws of thermodynamics in both equilibrium and 
non-equilibrium descriptions. We also show the relation 
between the two pictures. In Sec.~\uppercase\expandafter{\romannumeral 3}, 
we explore  the cases in the Einstein frame. 
Finally, we conclude in section \uppercase\expandafter{\romannumeral 4}.

%%%%%%%%%%%%%%%%%%%%%%%%%%%%%%%%%%%%%%%%%%%%%%%%%%%%%%%%%%%%%%%%%%%%%%%%%%%%%%%%%%%%%%%%%%%%%%%%
%-----------------------------------------------------------------------%
% Sec. II - Thermodynamics in Jordan Frame
%-----------------------------------------------------------------------%
\section{Thermodynamics in Jordan Frame}

The action of $f(R)$ gravity with matter is given by
\begin{equation}\label{A:FR}%%%%%%%%%%%
S = \frac{1}{2\kappa}\int \mathrm{d}^4x \sqrt{-g}\, f(R) + \sum_i S^{(i)}_{\text{M}}\,,
\end{equation}
where $\kappa\equiv8\pi G$ with $G$ the Newton's constant,
$f(R)$ is a function of the scalar curvature $R$, and 
$S^{(i)}_{\text{M}}=\int \mathrm{d}^4x\, \la_i$
correspond to the non-interacting matter actions with 
$i=(\text{m}$,$\text{r})$, representing
non-relativistic matter and radiation, respectively.
We use the disformal transformation of (\ref{disformal Def.})
in $f(R)$ gravity,
with $X := -(1/2)g^{\m\n}\pa_\m\f\pa_\n\f$ the kinetic term 
of the $\phi$ field. According to the disformal transformation
by Bekenstein, the matter fields directly couple to the 
background field $\gamma_{\mu\nu}$, while the matter action 
must be identified as 
$S^{(i)}_{\text{M}}[\g_{\m\n}]$, 
in which the model can be regarded as a special kind of the 
bimetric theory~\cite{Magueijo:2010zc}.
In this work, we will concentrate on the simple case with an 
assumption that matter is described on the Minkowski spacetime 
with metric given by
\begin{align}\label{MinkmatterMetric}
\eta_{\mu\nu} = A(\f,X)g_{\m\n} + B(\f,X)\pa_\m\f\pa_\n\f\,,
\end{align}
where $\eta_{\m\n} = \text{diag} (-1,+1,+1,+1)$ in our notation.
To examine how the disformal transformation will affect the
$f(R)$ gravity theory in cosmology, we assume our 
%{\color{red}
space to be
homogeneous and isotropic. As a result, the disformal field, 
$\phi$, only depends on time now. 
More precisely, the equation
$\pa_{\mu}\phi=(\dot{\phi},0,0,0)$ holds at present.
Subsequently, one can obtain $\d(\f,_\rho)$ in terms of $\d\gumn$ 
and $\d \f$, given by
\begin{align}
\d(\pa_{\beta}\f)= \bar{V}_\beta A_{\m\n} \delta g^{\m\n}
+\bar{V}_{\beta} \delta \f\,, \label{keyeqn01}
\end{align}
where
\begin{align}
\bar{V}_{\b} &= (\pa_{\b}\f)\frac{4\pa_{\f}A-2X\pa_{\f}B}
                  {(2X)(-4A,_X+2B+2XB,_X)}\,, \nonumber\\
A_{\m\n}     &= \frac{-Ag_{\m\n}-2A,_X\pa_\m\f \pa_\n\f 
                + XB,_X\pa_\m\f \pa_\n\f}{4\pa_{\f}A-2X\pa_{\f}B}\,.
\end{align}

%-----------------------------------------------------------------------%
The equations of motion from  the variations with respect to
$g^{\mu\nu}$ and $\phi$ are % and {\color{red}$\Psi_i$}
given by
\begin{align}%%%%%%%%%%%
FG_{\mu\nu} &= \kappa\sum_i\Big(T^{(i)}_{\m\n} + t^{(i)}_{\m\n}\Big) 
               + \kappa\hat{T}^{(\text{d})}_{\m\n}\,, \label{eom_fr01} \\
0           &= \frac{\pa\la_i}{\pa\f} 
               + \frac{(\pa_\a\f)(4\pa_\f A-2X\pa_\f B)}{(2X)(-4A,_X+2B+2XB,_X)}
                 \frac{\pa\la_i}{\pa (\pa_\a\f)}\,, \label{eom_phi01}
\end{align}
where $F:=d{f(R)}/d{R}$
and the comma denotes as the partial derivative.
The terms in the right-handed side of (\ref{eom_fr01})
correspond to the energy-momentum tensors of matter, 
disformally induced matter, and effective dark energy, defined by
\begin{align}%%%%%%%%%%%
T^{(i)}_{\m\n} 
&= \frac{-2}{\sqrt{-g}}
   \frac{\delta\la_i}{\delta \gumn}\,,\nonumber\\
   %\label{tmn_T01}
t^{(i)}_{\m\n} 
&= \frac{-2}{\sqrt{-g}}
   \bigg(\frac{-Ag_{\m\n}-2A,_X\pa_\m\f \pa_\n\f+XB,_X\pa_\m\f \pa_\n\f}
     {(2X)(-4A,_X+2B+2XB,_X)}\bigg)(\pa_\a \f)
   \frac{\pa\la_i}{\pa (\pa_\a \f)}\,, \nonumber\\%\label{tmn_t01}
\hat{T}^{(\text{d})}_{\m\n} 
&= \frac{1}{8\pi G}\bigg(\frac{1}{2}\gmn(f(R) - FR) 
   + \na_\mu\na_\nu{F} - \gmn\Box{F}\bigg),\label{Tdmn}
\end{align}
which are all assumed to be perfect fluids, given by
\begin{align}%%%%%%%%%%%
T^{(i)\,}_{\m\n} 
   &= (\r_{i} + P_{i})u_\m u_{\n} + P_{i}\,g_{\m\n}\,, \nonumber\\%\label{E:matter energy sress}\\
t^{(i)\,}_{\m\n}  
   &= (\r^{\text{(in)}}_{i} + P^{\text{(in)}}_{i})u_\m u_\n 
      + P^{\text{(in)}}_{i}g_{\m\n}\,,\nonumber\\%\label{E:induced energy sress}\\
\hat{T}^{\text{(d)}\,}_{\m\n}
   &= (\hat{\r}_{\text{d}} + \hat{P}_{\text{d}})u_\m u_\n 
      + \hat{P}_{\text{d}}\,g_{\m\n}\,, \label{E:matter energy sress}
\end{align}
where $\square F:=(1/\sqrt{-g})
\partial_{\mu}(\sqrt{-g} g^{\mu\nu}\partial_{\nu}F)$, $u^\m = dx^\m/d\tau$
is the unit four-vector with $\gumn u_\m u_\n = -1$,
$\r_{i}$ $(\r^{\text{(in)}}_{i})$, $\hat{\r}_{\text{d}}$
are the energy density
for the $i$th content of matter (induced matter) and dark energy, and 
$P_{i}$ $(P^{\text{(in)}}_{i})$, and $\hat{P}_{\text{d}}$ are the corresponding pressures, respectively.
%{\color{red}Since $\la_i$ is a function of $\eumn$, 
%we can apply chain rule so that the function 
%$\delta\la_i/\delta\eumn$ is included in each term.}

%-----------------------------------------------------------------------%
In this article, we will only focus on the 
\emph{flat} FLRW universe ($k=0$) in (\ref{A:FR}).
The corresponding metric is given by
\begin{equation}\label{FLRW}
ds^2 = -dt^2 + a^2(t)
\big( dx^2 + dy^2 + dz^2 \big)\,,
\end{equation}
with $a(t)$ the scale factor.
By comparing the components of $\eta_{\mu\nu}$, the relations
\begin{subequations}
\begin{align}%%%%%%%%%%%
A(\phi,X) &= 2XB(\phi,X)+1\label{pphi00}\,,\\
A(\phi,X) &=\frac{1}{a^2(t)} \label{pphi01}
\end{align}
\end{subequations}
can be derived, where $X=\dot{\phi}^2/2$ with the dot denoting
the derivative with respect to t.
As a result, the induced energy-momentum tensor in (\ref{Tdmn}) can be read as
%{\color{red}
\begin{align}%%%%%%%%%%%
t^{(i)}_{\m\n} 
= \frac{1}{a^3}\frac{g_{\m\n}a \dot{\phi}\ddot{\phi}
    -[3\dot{a}+(a^2 - 1)a \ddot{\phi} \dot{\phi}^{-1}]
    (\pa_\m \phi)(\pa_\n \phi)}{3 \dot{a} \dot{\phi}}
\bigg(\frac{\pa \la_i}{\pa \dot{\phi}}\bigg)\,.\label{tmn00}
\end{align}
%}
By taking the trace of (\ref{tmn00}), one gets
%{\color{red}
\begin{align}
3P^{\text{(in)}}_{i}-\rho^{\text{(in)}}_{i} 
= \frac{3 a \ddot{\phi} + 3\dot{a} \dot{\phi} 
    + a^3 \ddot{\phi}}{3 a^3 \dot{a}}
  \bigg(\frac{\pa \la_i}{\pa \dot{\phi}}\bigg)\,.
\end{align}
%}
We assume that the matter LaGrangian $\mathcal{L}_{i}$ 
is independent of the derivative of metric tensor,
leading to the fact that
\begin{align}
\frac{\delta \mathcal{L}_{i}}{\delta g^{\mu\nu}}
= \frac{\partial \mathcal{L}_{i}}{\partial g^{\mu\nu}}\,.
\end{align}
One also finds that
%{\color{red}
\begin{align}
\frac{\pa \la_i}{\pa \f}
= \frac{\pa \la_i}{\pa g_{\mu\nu}}\frac{\pa g_{\mu\nu}}{\pa \f}
=\bigg[ \frac{a^2 \dot{a} \dot{\phi} -a^2\dot{a} \dot{\phi} 
  -a^3\ddot{\phi}(1-a^2)}{\dot{\phi}^2}\bigg]\rho_{i}
  + \bigg(\frac{6a^2\dot{a}}{\dot{\phi}}\bigg)P_{i}\,,
\end{align}
%}
along with the equation of motion in (\ref{eom_phi01}) to be
%{\color{red}
\begin{align}%%%%%%%%%%%
\bigg[\frac{a^2 \dot{a} \dot{\phi} -a^2\dot{a} \dot{\phi}
    -a^3\ddot{\phi}(1-a^2)}{\dot{\phi}}\bigg]\rho_{i}
+ 6a^2\dot{a} P_{i}
+ \frac{a^3[a \dot{\phi}^{-1}\ddot{\phi}^2 (1-a^2) 
    - 3 \dot{a}\ddot{\phi}]}{3 a \ddot{\phi}
+ 3\dot{a} \dot{\phi}+a^3 \ddot{\phi}}
  (3P^{\text{(in)}}_{i}-\rho^{\text{(in)}}_{i}) = 0\,. \label{EOM for phi}
\end{align}
%}

%-----------------------------------------------------------------------%
Since the energy-momentum tensors of matter and induced matter
in (\ref{Tdmn}) all contain the derivative terms in the matter LaGrangian $\la_i$, 
we can connect matter and induced matter and 
derive the relations of energy densities and pressures 
between them.
From the  $tt$-components in (\ref{Tdmn}) and (\ref{E:matter energy sress}), 
the induced energy density can be expressed as
\begin{align}
\r_{i}^{(\text{in})} = \lambda\,\r_{i}\label{rhophi01}
\end{align}
with 
\begin{align}\label{Eq:Disformal coupling}
\lambda = \frac{(1-a^2)(3 \dot{a} \dot{\phi} + a^3 \ddot{\phi})}{3 \dot{a} \dot{\phi}}\,,
\end{align}
which is related to the disformal field $\phi$ as expected.
Similarly,
one can write the induced pressure in terms of the energy density by
\begin{align}
P^{(\text{in})}_{i} 
= -\frac{a \ddot{\phi}(1-a^2)}
    {3 \dot{a} \dot{\phi}}\rho_i\,.\label{induced pressure}
\end{align}
In addition, the equations of state (EoS) of matter and 
induced matter are given by 
$w_i := P_i/\r_i$ and $w_{i}^{(\text{in})} 
  := P_{i}^{(\text{in})}/\r_{i}^{(\text{in})}$, respectively.

%-----------------------------------------------------------------------%
By using (\ref{rhophi01}) and (\ref{induced pressure}), 
one can describe $P^{(\text{in})}_i$ in terms of $\rho_i$
\begin{align}%%%%%%%%%%%
P_{i}^{(\text{in})}  &=w_{i}^{(\text{in})} \lambda \r_i\,, \label{pphi01}
\end{align}
with  the induced EoS of $w_{i}^{(\text{in})}$, given by
\begin{align}\label{Eq:Induced EoS}
w_{i}^{(\text{in})} = - \frac{a \ddot{\phi}}{3 \dot{a} \dot{\phi} + a^3 \ddot{\phi}} 
\equiv w^{(\text{in})}\,,
\end{align}
independent of  the ordinary matter content.
Please note that (\ref{pphi01}) can also be presented as 
$P^{(\text{in})}_i=(w^{(\text{in})}/w_i) \lambda P_{i}\,$. 
The induced matter can be totally written in terms of ordinary matters, implying the modification of the matter contents of 
the universe through the disformal coupling.

%-----------------------------------------------------------------------%
The reason that the induced matter content is related by 
the ordinary matter content in (\ref{rhophi01}) and~(\ref{pphi01}) 
can be understood in a sense that the disformal field, $\phi$, 
is always generated to offset the gravitation field, $g_{\m\n}$, 
to force the physical metric, which governs the equation of motion 
of ordinary matter~\cite{Bekenstein:1992pj}, to Minkowski one. 
Therefore, once the ordinary energy and pressure, $\rho_i$ and $P_i$, 
are given, one can obtain 
 (\ref{Eq:Disformal coupling}) and (\ref{Eq:Induced EoS}), 
so that $\r_{i}^{(\text{in})}$ and $P_{i}^{(\text{in})}$ 
can be found by (\ref{rhophi01}) and (\ref{pphi01}), respectively.

%-----------------------------------------------------------------------%
We assume that there contain non-relativistic matter(m) and 
radiation(r) in our model.
For non-relativistic matter, 
the pressure is zero, $P_{\text{m}} = 0$, 
which gives the vanished matter EoS of $w_\text{m}=0$, while 
 $w_{\text{r}} = 1/3$ for radiation.
The resulting induced energy density and pressure are given by
\begin{equation}\label{E: induce m}
\r_{\text{m}}^{(\text{in})} = \lambda \r_\text{m} \,, 
\quad P_{\text{m}}^{(\text{in})}  = \lambda\, w^{(\text{in})} \, \r_\text{m}\,.
\end{equation}
Similarly, one has
\begin{equation}\label{E: induce r}
\r_{\text{r}}^{(\text{in})} = \lambda\, \r_\text{r}\,, 
\quad P_{\text{r}}^{(\text{in})}  =\lambda\,w^{(\text{in})}\,  \r_{\text{r}}\,.
\end{equation}
%{\color{blue}Since the pressure of radiation is not zero, this ensures that the disformal coupling does give the alternative contribution.
%We could also find that the disformal field does not effect the gravity, and that the part of matter ends up no contribution.}
According to (\ref{E: induce m}), (\ref{E: induce r}) and 
(\ref{FLRW}), we obtain the modified Friedmann equations as
\begin{align}%%%%%%%%%%%
3FH^2      &= 8\pi G\Big((1 + \l )\bar{\rho}_{\text{M}}\Big)
              - 3H\dot{F} - \frac{1}{2}(f - FR),\nonumber\\
-2F\dot{H} &= 8\pi G\Big(\bar{\rho}_{\text{M}}+P_r
              + \l \big(1+w^{(\text{in})}\big)\bar{\rho}_{\text{M}}\Big)
              + \ddot{F} - H\dot{F}\label{eomfrhdot_01}\,.
\end{align}
where $\bar{\rho}_{\text{M}} = \r_\text{m}+\r_\text{r}$.

%{\color{red}
By simply rearranging the $H^2$ and $\dot{H}$ terms to the left-hand side (LHS) and  the
	others to the right-hand side (RHS) in  (\ref{eomfrhdot_01}), one obtains that
%One can rewrite (\ref{eomfrhdot_01}) into the following forms
%	
	\begin{align}%%%%%%%%%%%
	H^2     &= \frac{8\pi G_{\text{eff}}}{3}
	\bigg(\bar{\rho}_{\text{M}} +\hat{\r}_{\text{d}} 
	+ \l\bar{\rho}_{\text{M}}\bigg)\,,\nonumber\\
	\dot{H} &= -4\pi G_{\text{eff}}
	\bigg(\big(\bar{\rho}_{\text{M}} 
	+ \hat{\r}_{\text{d}} + P_{\text{r}} + \hat{P}_{\text{d}}\big) 
	+ \l\big(1+w^{(\text{in})}\big)\bar{\rho}_{\text{M}}\bigg)\,,\label{noneq01}
	\end{align}
	where
	\begin{align}%%%%%%%%%%%
	\hat{\r}_{\text{d}} &= \frac{1}{8\pi G}
	\bigg(\frac{1}{2}(FR - f) - 3H\dot{F}\bigg)\,,\nonumber\\
	\hat{P}_{\text{d}} &= \frac{1}{8\pi G}
	\bigg( \ddot{F} +2H\dot{F} - \frac{1}{2}(FR - f)\bigg) \label{noneqDE},
	\end{align}
	where $G_{\text{eff}}=G/F$.  The subscript ``d''
	in \eqref{noneq01} represents the effective dark energy component 
	of $f(R)$  in the non-equilibrium picture.
	In addition, one can express (\ref{eomfrhdot_01}) in terms of 
	the following by adding $H^2$ and $\dot{H}$ terms on the both sides
	of the two equations, respectively, given by
	\begin{align}%%%%%%%%%%%
	H^2     &= \frac{8\pi G}{3}\bigg((\bar{\rho}_\text{M}+\r_{\text{d}}) 
	+\l\bar{\rho}_\text{M}\bigg),\nonumber\\%\label{eomeqph2_01}\\
	\dot{H} &= -4\pi G\bigg(\big(\bar{\rho}_\text{M}
	+\r_{\text{d}} + P_{\text{r}} + P_{\text{d}}\big) 
	+ \l\big(1+ w^{\text{(in)}}\big)\bar{\rho}_\text{M}\bigg)\,,\label{eq01}
	\end{align}
	where 
	\begin{align}%%%%%%%%%%%
	\r_\text{d} &= \frac{1}{8\pi G}\bigg[ \frac{1}{2}(FR - f) 
	- 3H\dot{F} + 3H^2(1-F)\bigg],\nonumber\\
	P_\text{d}  &= \frac{1}{8\pi G}\bigg[ \ddot{F} +2H\dot{F} 
	- \frac{1}{2}(FR - f) - (1-F)(2\dot{H} + 3H^2 )\bigg]\,.\label{eqDE}
	\end{align}
	with $G$ the ordinary Newtonian constant.
	It will be shown in the later context that (\ref{noneq01}) corresponds 
	to the non-equilibrium description of thermodynamics, whereas (\ref{eq01})
	the equilibrium one.
	%}

%------------------------------------------------------------------%
%Subsec. A - Non-equilibrium Description of Thermodynamics in f(R) Gravity 
%------------------------------------------------------------------%
\subsection{Non-Equilibrium Description of Thermodynamics in $f(R)$ Gravity }

%-------------------------------------------------------------%
%Subsubsec. 1 - First Law in Non-equilibrium Description
%-------------------------------------------------------------%
\subsubsection{First Law in Non-Equilibrium Description}

To study thermodynamics, we start with the non-equilibrium picture.
{color{red}According to \eqref{noneq01} and \eqref{noneqDE}},
it is clear that the extra terms in the RHSs of (\ref{noneq01}) arise
from the assumption that the matter metric is related to the gravitational one
through the disformal transformation.
We can put Friedmann equations (\ref{noneq01}) in more compact forms,
\begin{align}
H^2 &=  \frac{8\pi G}{3F} \hat{\r}_t \,, \nonumber\\
\dot{H} &= -\frac{4 \pi G}{F} (\hat{\r}_t + \hat{P}_t)\,.
\end{align}
where 
\begin{align}
\hat{\r}_t &= \rho_{\text{M}}+\hat{\r}_{\text{d}} 
            = \bar{\rho}_{\text{M}} + \l \bar{\rho}_{\text{M}} 
              + \hat{\r}_{\text{d}}\,,  \nonumber\\
\hat{P}_t &= P_{\text{M}} + \hat{P}_{\text{d}} 
           = P_{\text{r}} + \l w^{\text{(in)}} \bar{\r}_{\text{M}} 
             + \hat{P_{\text{d}}}\,.
\end{align}
Here, $\rho_{\text{M}}$ is represented as the total matter density
including  induced matter.

%-----------------------------------------------------------------------%
Please note that the energy-momentum tensors from matter and 
radiation obey 
$\na_\m({T}^{(\text{m})}{}^\m{_\n} + {T}^{(\text{r})}{}^\m{_\n}
 + t^{(\text{m})\m}{_\n} + t^{(\text{r})\m}{_\n}) =0$,
resulting in the continuity equation
\begin{align}%%%%%%%%%%%
\dot{\rho}_\text{M}+3H(\rho_\text{M}+P_\text{M}) = 0\label{contieq01}.
\end{align}
Furthermore, the continuity equation for dark energy  is found to be
\begin{align}%%%%%%%%%%%
\dot{\hat\r}_\text{d} + 3H(\hat{\r}_{\text{d}} + \hat{P}_{\text{d}}) 
= \frac{3H^2\dot{F}}{8\pi G}\,.
\end{align}
In the non-equilibrium picture, the theory requires $\dot{F} \neq 0$,
so that the above equation does not vanish.
%{\color{red}
This leads to the non-equilibrium
	description of thermodynamics in $f(R)$ gravity in the Jordan frame.
	%}

%-----------------------------------------------------------------------%
The four-dimensional FLRW metric in (\ref{FLRW}) can be rewritten as
\begin{align}%%%%%%%%%%%
ds^2 = h_{ab}dx^a dx^b + \rb^2 d\O^2,\label{flrw2}
\end{align}
where $h_{ab} = \text{diag}(-1, a^2(t))$ is the two-dimensional metric 
with $a$, $b=(0,1)$, while $x^0 = t$, and $x^1=r$.
To examine the thermodynamic properties in $f(R)$, 
%{\color{red}
we require  the apparent horizon $\rb_A$ to
satisfy the condition 
$h^{ab}(\pa_{a}\rb)(\pa_{b}\rb) = 0$~\cite{Cai:2005ra}.
In the FLRW spacetime, it is given by
$\rb_A = H^{-1}$, so that
$d\rb_A = - \rb_{A}^{2} \dot{H}dt$.
Combining with (\ref{noneq01}), we find %please confirm the ref. 
\begin{align}%%%%%%%%%%%
\frac{Fd\rb_A}{4\pi{G}}
= \rb_A^2\bigg(\big(\bar{\rho}_\text{M} + \hat{\r}_{\text{d}} 
    + P_{\text{r}} + \hat{P}_{\text{d}}\big) 
  + \l\big(1+w^{(\text{in})}\big)\bar{\rho}_\text{M}\bigg)dt\,.
\end{align}

%-----------------------------------------------------------------------%
We use the dynamical entropy 
$\hat{S} = A/(4G_{\text{eff}})$ in the $f(R)$ 
gravity theory at a given horizon, 
where $A = 4\pi\rb_A^2$ is the area of the apparent horizon 
and $G_{\text{eff}}$ is the effective Newton's constant.
The dynamical entropy was first discussed by Wald~\cite{Wald:1993nt}, who proposed that the entropy of the black hole in GR is a
Noether charge~\cite{Wald:1993nt,Iyer:1994ys}.
It was also shown that this entropy is 
associated with the effective gravitational coupling,
which depends on the variation of gravity LaGrangian 
with respect to the Riemann curvature tensor~\cite{Brustein:2007jj}.
For $f(R)$ gravity, in the non-equilibrium picture, 
since $G_{\text{eff}} = G/F$ is the effective gravitational 
coupling, we have the entropy 
\cite{Briscese:2007cd,Bamba:2010kf,Jacobson:1993vj,Iyer:1994ys,Cognola:2005de}
\begin{align}
\hat{S} = \frac{FA}{4G}\,.
\end{align}

%-----------------------------------------------------------------------%
The associated temperature at the apparent horizon is proportional to
the surface gravity $\k_s$, given by \cite{Bardeen:1973gs}
\begin{align}
T = \frac{|\k_s|}{2\pi}\,.
\end{align}
By substituting $\kappa_s = 1/(2\sqrt{-h})\pa_\a
 \big( \sqrt{-h}h^{\a\b}\pa_\b\rb\big)|_{\rb=\rb_A}$
 in the above equation, we get 
\begin{align}
T = \frac{1}{2\pi\rb_A}\bigg(1-\frac{\dot\rb_A}{2H\rb_A}\bigg)\,,
\end{align}
which should be positive~\cite{Bamba:2010kf}.
As a result, $Td\hat{S}$ can be written as
\begin{align}%%%%%%%%%%%
Td\hat{S} = 4\pi \rb_A^2\Big((\bar{\rho}_\text{M} 
            + \hat{\r}_{\text{d}} + P_{\text{r}} + \hat{P}_{\text{d}})
            + \l(1+w^{(\text{in})})\bar{\rho}_\text{M}\Big)
              \Big(1+\frac{\dot{H}\rb_A^2 }{2}\Big)dt 
            + \frac{T}{G}\pi\rb_A^2dF\,.  \label{tempdS}
\end{align}
The Misner-Sharp energy within the apparent horizon
in $f(R)$ gravity is given by 
\cite{Wu:2007se,Wu:2008ir,Bamba:2010kf,Gong:2007md}
\begin{align}%%%%%%%%%%%
\hat{E} = \frac{\rb_A}{2G_{\text{eff}}}\,,
\end{align}
which can be regarded as the total energy within the 
apparent horizon, i.e., 
$ \hat{E}=V\hat{\rho}_t$, 
where $V = 4\pi\rb_A^3/3$ is the volume of the horizon.

%-----------------------------------------------------------------------%
Differentiating the Misner-Sharp energy, we have
\begin{align}%%%%%%%%%%%
d\hat{E} 
&= 4 \pi \rb_A^2 
   \big(\bar{\rho}_\text{M} 
     +\hat{\r}_{\text{d}} +\l\bar{\rho}_\text{M} \big) 
   d\rb_A+\frac{\rb_A}{2G}dF \nonumber\\
& \quad-4 \pi H \rb_A^3\Big(\big(\bar{\rho}_\text{M} 
     + \hat{\r}_{\text{d}} + P_{\text{r}} + \hat{P}_{\text{d}}\big) 
   + \l\big(1+w^{(\text{in})}\big)\bar{\rho}_\text{M}\Big)dt\,. \label{eqndE}
\end{align}
By introducing the work density \cite{Cai:2005ra,Bamba:2010kf}
$\hat{W} = -\frac{1}{2} \big(T^{(\text{m})ab} + T^{(\text{r})ab} 
           + t^{(\text{m})ab} + t^{(\text{r})ab} 
           + \hat{T}^{(\text{d})ab}\big)h_{ab}$,
with the perfect fluids, we obtain
\begin{align}
\hat{W} = \frac{1}{2} \Big(\big(\bar{\rho}_\text{M} + \hat{\r}_{\text{d}} 
          - P_{\text{r}} - \hat{P}_{\text{d}}\big) 
          + \l\big(1-w^{(\text{in})}\big)\bar{\rho}_\text{M}\Big)
\end{align}
and the new quantity
\begin{align}%%%%%%%%%%%
d_i\hat{S} &= -\frac{1}{T}\frac{\rb_A}{2G}(1 + 2\pi\rb_A T)dF\,,
\end{align}
we obtain
\begin{align}%%%%%%%%%%%
Td\hat{S} + Td_i\hat{S} = -d\hat{E} + \hat{W}dV\,,\label{neq1stlaw}
\end{align}
which is the first law in the non-equilibrium picture.
We can see that it is similar to the result in 
$f(R)$ gravity~\cite{Akbar:2006mq,Bamba:2009ay}. 
The first law of thermodynamics is still valid
when we introduce a disformal relation in (\ref{MinkmatterMetric}).

%-------------------------------------------------------------%
%Subsubsec. 2 - Second Law in Non-equilibrium Description
%-------------------------------------------------------------%
\subsubsection{Second Law in Non-Equilibrium Description}

To describe the second law of thermodynamics in the non-equilibrium picture, 
we write down the Gibbs function in terms of the total energy 
$\hat{\r}_t$ pressure $\hat{P}_t$, given by
\begin{align}%%%%%%%%%%%
Td\hat{S}_t = d(\hat\r_t V) + \hat{P}_tdV 
= Vd\hat\r_t + (\hat\r_t + \hat{P}_t)dV\,.
\end{align}
The time derivatives of entropies lead to
\begin{align}\label{dS/dt}
\frac{d\hat{S}}{dt} + \frac{d_i\hat{S}}{dt} + \frac{d\hat{S}_t}{dt} 
= -\frac{48\pi^2 \dot{H}}{R H^3}\bigg((\bar{\rho}_\text{M} 
  + \hat{\r}_{\text{d}} + P_{\text{r}} + \hat{P}_{\text{d}}) 
  + \l\big(1 + w^{(\text{in})}\big)\bar{\rho}_\text{M}\bigg)\,,
\end{align}
where the relations of 
$R = 6(\dot{H} + 2H^2)$ and $T^{-1}=24 \pi H/R$ 
have been used. It can be seen that the influence of 
the disformal transformation on the rate of the entropy 
change is the corrections of the energy densities and pressures.
By using the Friedmann equations (\ref{noneq01}), %please confirm the ref. 
(\ref{dS/dt}) becomes 
\begin{align}\label{Eq:entropy rate in j-frame}
\frac{d\hat{S}}{dt} + \frac{d_i\hat{S}}{dt} + \frac{d\hat{S}_t}{dt} 
= \frac{12 \pi F \dot{H}^2}{GR H^3}\,.
\end{align}
Please note that for our current accelerating expansion of the universe, 
the Hubble parameter $H$ and Ricci scalar $R$ are positive.
%{\color{red}
On the other hand, to avoid the ghost  and instability problems, 
the positive condition of $F=df/dR$ and $d^2f/dR^2$ should be satisfied.
It can be checked that the equation (\ref{Eq:entropy rate in j-frame}) 
is positive for all viable $f(R)$ gravity theories
due to the viable condition of
$F=df/dR>0$ \cite{Nunez:2004ji,Faraoni:2007yn}.
%}
As a result, (\ref{Eq:entropy rate in j-frame}) is consistent with 
the second law of thermodynamics, i.e.
\begin{align}%%%%%%%%%%%
\frac{d\hat{S}}{dt} + \frac{d_i\hat{S}}{dt} + \frac{d\hat{S}_t}{dt}
\geq 0\,, \label{2ndlawneq}
\end{align}
when considering the disformal transformation
with the Minkowski matter metric. This conclusion  is the same as 
that in $f(R)$ gravity without the disformal transformation~\cite{Bamba:2009ay}.
It is clear that our results in (\ref{neq1stlaw}) and (\ref{2ndlawneq}) 
are reduced to those in $f(R)$ without the disformal transformation 
in the non-equilibrium picture in the limit of $\l \rightarrow 0$.

%------------------------------------------------------------------%
%Subsec. B - Equilibrium Description of Thermodynamics in f(R) Gravity 
%------------------------------------------------------------------%
\subsection{Equilibrium Description of Thermodynamics in $f(R)$ Gravity}

%-------------------------------------------------------------%
%Subsubsec. 1 - First Law in Equilibrium Description
%-------------------------------------------------------------%
\subsubsection{First Law in Equilibrium Description}

In the equilibrium picture, we use (\ref{eq01}) and
\eqref{eqDE} as modified Friedmann equations.
The energy-momentum tensor of dark energy is given by
\begin{align}%%%%%%%%%%%
T^{(\text{d})}_{\m\n} 
= \frac{1}{8\pi G}\Big(\frac{1}{2}\gmn(f(R) 
  - FR) +\na_\mu\na_\nu{F} - \gmn\Box{F} - (1-F)G_{\m\n}\Big)\label{tdmn}\,.
\end{align} 
Since dark energy in the equilibrium picture is treated as 
a perfect fluid, $T^{(\text{d})}_{\m\n}$ can be written in the form of
\begin{align}
T^{(\text{d})}_{\m\n}
&= (\r_{\text{d}} + P_{\text{d}})u_\m u_\n + P_{\text{d}}\,g_{\m\n}\,.
\end{align}
Consequently, the modified Einstein equation in (\ref{eom_fr01}) becomes
\begin{align}%%%%%%%%%%%
G_{\m\n} = 8 \pi G\Big(T^{(\text{m})}_{\m\n} + T^{(\text{r})}_{\m\n} 
           + t^{\text{(m)}}_{\m\n} + t^{\text{(r)}}_{\m\n} 
           + T^{(\text{d})}_{\m\n} \Big)\,. \label{Einsteineq_equil}
\end{align}

%-----------------------------------------------------------------------%
Since the left-handed side of (\ref{Einsteineq_equil}) has the property 
of $\na_\mu G^{\mu\nu}=0$, the total energy-momentum 
should obey the continuity equation 
$\na_\mu(T^{(\text{m})}{}^\m{}_\n+T^{(\text{r})}{}^\m{}_\n 
  + t^{(\text{m})}{}^{\m}{}_{\n} + t^{(\text{r})}{}^{\m}{}_{\n} 
  + T^{(\text{d})}{}^\m{}_\n) = 0$.
After applying the FLRW metric to the continuity equation
\begin{align}%%%%%%%%%%%
\dot\r_t + 3H(\r_t + P_t) = 0,
\end{align}
with 
$\rho_t = (\bar{\rho}_\text{M}+\r_{\text{d}}) +\l\bar{\rho}_\text{M}$ 
and
$P_t = P_{\text{r}} + P_{\text{d}} + \l w^{(\text{in})}\bar{\rho}_\text{M}$,
one finds that dark energy also obeys 
its own continuity equation
\begin{align}
 \dot\r_\text{d} + 3H(\r_\text{d} + P_\text{d})= 0\,,
\end{align}
along with the one for matter and radiation, given by
\begin{align}
\na_\m\big(T^{(\text{m})}{}^{\mu}{_\nu} + T^{(\text{r})}{}^{\mu}{_\nu}
    + t^{(\text{m})}{}^{\mu}{_\nu} + t^{(\text{r})}{}^{\mu}{_\nu}\big) = 0\,,
\end{align}
resulting in
\begin{align}
\na_\mu\big(T^{(\text{m})}{}^\m{}_\n+T^{(\text{r})}{}^\m{}_\n 
+ t^{(\text{m})}{}^{\m}{}_{\n} + t^{(\text{r})}{}^{\m}{}_{\n} 
+ T^{(\text{d})}{}^\m{}_\n\big) 
= 0
\end{align}
for the total energy-momentum. 
Hence, the equilibrium description of thermodynamics can be advocated.
Using the Bekenstein-Hawking entropy $S=A/4G$ and 
Hawking temperature $T=|\kappa_s|/2\pi$, we have
\begin{align}%%%%%%%%%%%
TdS = 4\pi\rb_A^2
      \bigg(\big(\bar{\r}_\text{M}
        + \r_{\text{d}} + P_{\text{r}} + P_{\text{d}}\big) 
        + \l\big(1 + w^{(\text{in})}\big)\bar{\r}_\text{M}\bigg)
      \bigg(1+\frac{\dot{H}\rb_A^2 }{2}\bigg)dt\,. \label{1steqp01}
\end{align}
Defining the Misner-Sharp energy within the apparent horizon as 
$E = \rb_A/2G = V\rho_t$, 
we obtain
\begin{align}%%%%%%%%%%%
dE &= 4\pi\rb_A^2\bigg((\bar{\r}_{\text{M}}+\r_{\text{d}}) 
      + \l\bar{\r}_{\text{M}}\bigg)d\rb_A \nonumber\\ 
   &\quad -4\pi\rb_A^2\bigg(\big(\bar{\rho}_\text{M}+\r_{\text{d}} 
          + P_{\text{r}} + P_{\text{d}}\big) 
          + \l\big(1+ w^{\text{(in)}}\big)\bar{\rho}_\text{M}\bigg)dt\,. \label{MSEeqp01}
\end{align}
With (\ref{1steqp01}) and (\ref{MSEeqp01}), we derive
\begin{align}%%%%%%%%%%%
TdS = -dE + WdV\,.\label{1stlaweqp}
\end{align}
where $W$ is the work density, defined by
\begin{align}
W &= -\frac{1}{2} 
     \bigg(T^{(\text{m})ab} + T^{(\text{r})ab} 
       + t^{(\text{m})ab} + t^{(\text{r})ab} 
       + T^{(\text{d})ab}\bigg)h_{ab} \nonumber\\
  &= \frac{1}{2} 
     \bigg(\big(\bar{\rho}_\text{M} + \r_{\text{d}} 
       - P_{\text{r}} - P_{\text{d}}\big)
       + \l\big(1 - w^{\text{(in)}}\big)\bar{\rho}_\text{M}\bigg)\,.
\end{align}
Please note that (\ref{1stlaweqp}) is the first law of thermodynamics in the equilibrium picture of $f(R)$ gravity,
in which there is no other entropy contribution.
It is because dark energy satisfies its own continuity equation in the
equilibrium picture, whereas the non-equilibrium one does not.

%-------------------------------------------------------------%
%Subsubsec. 2 - Second Law in Equilibrium Description
%-------------------------------------------------------------%
\subsubsection{Second Law in Equilibrium Description}

To describe the second law in the equilibrium picture, 
we take the Gibbs function in terms of the total energy 
$\r_t$ and pressure $P_t$, given by
\begin{align}%%%%%%%%%%%
TdS_t = d(\r_t V) + P_tdV = Vd\r_t + (\r_t + P_t)dV\,.
\end{align}
The time derivatives of the entropies can be written as
\begin{align}%%%%%%%%%%%
\frac{d}{dt}(S + S_t) 
= -\frac{48 \pi^2 \dot{H}}{R H^3}
  \bigg((\bar{\rho}_\text{M}+\r_{\text{d}} 
    + P_{\text{r}} + P_{\text{d}})
    + \l(1+ w^{\text{(in)}})\bar{\rho}_\text{M}\bigg)\,.
\end{align}
By using (\ref{eomeqphdot_01}),
we derive 
\begin{align}%%%%%%%%%%%
\frac{d}{dt}(S + S_t) 
= \frac{12\pi\dot{H}^2}{GRH^3}\geq 0\,,\label{2ndlaweq}
\end{align}
which is the second law of thermodynamics.
We see that the difference between the two descriptions is related to the function $F$.
In the limit $\l \rightarrow 0$, the results in (\ref{1stlaweqp}) and (\ref{2ndlaweq}) 
can be also reduced to those in $f(R)$ gravity without the
disformal transformation in the equilibrium picture.

%-------------------------------------------------------------%
%Subsubsec. 3 - Relatin Between  Equilibrium And Non-equilibrium Description
%-------------------------------------------------------------%
\subsubsection{Entropy Difference Relation of Equilibrium and Non-Equilibrium Descriptions}

Comparing the definitions of the effective
dark energy in two frames (\ref{Tdmn}) and (\ref{tdmn}), 
we find that
\begin{align}%%%%%%%%%%%
T^{(d)}_{\m\n} = \hat{T}^{(d)}_{\m\n} - \frac{1 - F}{8 \pi G}G_{\m\n}\,.
\end{align}
As a result, we relate the entropies in the two descriptions as
\begin{align}%%%%%%%%%%%
dS = d\hat{S} + d_i\hat{S} + \frac{\rb_A}{2GT} 
     - \frac{2\pi\rb_A^4}{G}H\dot{H}(1-F)dt\,.
\end{align}
After some calculations, we obtain
\begin{align}%%%%%%%%%%%
dS = \frac{1}{F}d\hat{S} + \frac{1}{F}
     \frac{2H^2 + \dot{H}}{4H^2 + \dot{H}}d_i\hat{S}\,.\label{eqvsnoneq}
\end{align}

%%%%%%%%%%%%%%%%%%%%%%%%%%%%%%%%%%%%%%%%%%%%%%%%%%%%%%%%%%%%%%%%%%%%%%%%%%%%%%%%%%%%%%%%%%%%%%%%
%-----------------------------------------------------------------------%
% Sec. III - Thermodynamics in Einstein Frame
%-----------------------------------------------------------------------%
\section{Thermodynamics in Einstein Frame}

%-----------------------------------------------------------------------%
To discuss thermodynamics in the Einstein frame, 
we apply the conformal transformation to $f(R)$ gravity.
The metric tensor transforms as
$\ti{g}_{\a\b}(x^\m) = \O^2(x^\m)g_{\a\b}(x^\m)$ which leads to 
$\eta_{\mu\nu}=\O^{-2}A\ti{g}_{\mu\nu}+B\pa_\m\phi\pa_\n\phi$.
Due to the conformal transformation, 
it can be found that the matter LaGrangian densities 
are non-minimally coupled by $\O$
through the disformal transformation.

%-----------------------------------------------------------------------%
The action in the Einstein frame can be achieved by
the constraint of $\O^{-2}F = 1$, which is read as
\begin{align}\label{actionEF01}%%%%%%%%%%%
S = \int \mathrm{d^4}x \sqrt{-\ti{g}}\bigg[\frac{1}{2\kappa}\ti{R}
    -\frac{1}{2}\ti{g}^{\m\n}\pa_\m \o \pa_\n \o - V(\o)\bigg]
 + \sum_i S^{(i)}_{\text{M}}[\o,\ti{g}_{\mu\nu},\phi, \Psi_{\text{M}}]\,,
\end{align}
where the conformal scalar is defined as
$\o = \a\ln\O$ with $\a=\sqrt{3/(4\pi{G})}$ 
and 
$V(\o)=(FR-f)/{2 \kappa F^2}$.
Since the thermodynamic behavior does not affected 
by the magnitude of time interval, 
we can define 
$d\ti{t} = \O\, dt$ and $\ti{a}(\ti{t}) = \O\, a(t)$ 
in the Einstein frame for simplicity.
Thus, the FLRW metric becomes~\cite{DeFelice:2010aj}
\begin{align}%%%%%%%%%%%
d\ti{s}^2 
= -d\ti{t}^2 + \ti{a}^2(\ti{t})\bigg(dr^2
  + r^2d\theta^2 + r^2\sin^2{\theta}d\varphi^2\bigg)\,.
\end{align}
Again, we have used the constraint $\d\emn = 0$ to give the relation of
\begin{align}
\d(\pa_{\beta}\f) 
&= (\pa_{\beta}\f)
   \bigg(\frac{4\pa_\f A-2X\pa_\f B}
          {(2X)(-4\pa_XA+2B+2X\pa_X B)}\bigg) 
   \delta \f \nonumber \\
&\quad + (\pa_{\beta}\f)
       \bigg(\frac{-\O^{-2}A\ti{g}_{\m\n}
         - 2\pa_XA\,\pa_\m\f\,\pa_\n\f
         + X\pa_XB\,\pa_\m \f\,\pa_\n \f}
         {(2X)(-4\pa_XA+2B+2X\pa_XB)}\bigg)
         e^{\frac{2\o}{\alpha}} \,\delta \tilde{g}^{\m\n} \nonumber \\
&\quad + \frac{2}{\alpha}
       \bigg(\frac{-\O^{-2}A\ti{g}_{\m\n}-2\pa_X A\pa_\m\f\,\pa_\n\f 
         + X\pa_X B\pa_\m \f\,\pa_\n\f}{(2X)(-4\pa_X A+2B+2X\pa_X B)}\bigg)
       g^{\m\n}(\pa_{\beta}\f) \, \delta\o\,.
\end{align}
Varying the action (\ref{actionEF01}) with respect to
$\ti{g}^{\mu\nu}$, $\phi$ and $\o$,
we obtain the field equations
\begin{align}%%%%%%%%%%%
\ti{G}_{\m\n} &= \kappa\bigg(\sum_i \Big(\ti{T}^{(i)}_{\m\n} 
                 + \ti{t}^{(i)}_{\m\n}\Big) 
                 + \ti\mft_{\m\n}\bigg),\nonumber \\%\label{eomEFgmn01}\\
0             &= \frac{\pa\la_{i}}{\pa\phi} + \frac{(\pa_{\beta}\f)(4\pa_\f A 
                 - 2X\pa_\f B)}{(2X)(-4\pa_XA+2B+2X\pa_X B)}
                 \frac{\pa\la_i}{\pa (\pa_\b \f)} ,\nonumber \\%\label{eomEFphi01}\\
0             &= \frac{1}{\sqrt{-\ti{g}}}\pa_\m\Big(\sqrt{-\ti{g}} \, 
                 \ti{g}^{\m\n} \pa_\n \o\Big) - \frac{\pa V}{\pa \o} 
                 + \sum_i\bigg(\frac{1}{\sqrt{-\ti{g}}}\frac{\delta\la_i}{\delta\o} 
                 - \frac{1}{\alpha}\,\ti{g}^{\m\n} \ti{t}^{(i)}_{\m\n}\bigg)\,, \label{eomEFomega01}
\end{align}
where the quantities 
$\ti{T}^{(i)}_{\m\n}$, $\ti{t}^{(i)}_{\m\n}$ and $\ti\mft_{\m\n}$ 
are the energy-momentum tensors for matter, induced matter
and the conformal scalar in Einstein frame, defined as
\begin{align}%%%%%%%%%%%
\ti{T}^{(i)}_{\m\n} 
&= \frac{-2}{\sqrt{-\ti{g}}}
   \frac{\delta\la_{i}}{\delta\tgumn}\,, \nonumber\\%\label{ti_Tmn01}\\
\ti{t}^{(i)}_{\m\n} 
&= \frac{-2}{\sqrt{-\ti{g}}}\frac{\pa \la_{i}}{\pa(\pa_\a\f)}
   \bigg(\frac{A\O^{-2}\ti{g}_{\m\n}+2\pa_X A\pa_\m\f\,\pa_\n\f-X\pa_X B\pa_\m \f\,\pa_\n\f}
     {(2X)(-4\pa_X A+2B+2X\pa_X B)}\bigg)
   e^{\frac{2\o}{\a}}\pa_{\a}\f\,, \nonumber\\%\label{ti_tmn01}\\
\ti\mft_{\m\n} 
&= \ti{g}_{\m\n}\bigg(-\frac{1}{2}\ti{g}^{\a\b}\pa_\a\o\pa_\b\o 
     - V(\o)\bigg) + \pa_\m\o\pa_\n\o\,,\label{ti_ttmn01}
\end{align}
respectively.
With the perfect fluid assumption,
the above equations in (\ref{ti_ttmn01}) are written as
\begin{align}%%%%%%%%%%%
\ti{T}^{(i)}_{\m\n} 
&= (\ti\r_{i} + \ti{P}_{i})\ti{u}_\m\ti{u}_\n 
    + \ti{P}_{i}\tgmn\,, \nonumber\\%\label{PFti_Tmn01}\\
\ti{t}^{(i)}_{\m\n} 
&= (\ti\r^{\text{(in)}}_{i} + \ti{P}^{\text{(in)}}_{i})\ti{u}_\m\ti{u}_\n 
    + \ti{P}^{\text{(in)}}_{i}\,\tgmn\,, \nonumber\\ %\label{E:con induced emt},  \\
\ti\mft_{\m\n} 
&= (\ti\r_\o + \ti{P}_\o)\ti{u}_\m\ti{u}_\n + \ti{P}_\o\tgmn\,, \label{PFti_ttmn01}
\end{align}
respectively, 
where $\ti{u}^\m=d x^\m/d \ti{\tau}$
is defined with $\ti{\tau}$ the conformal proper time.

%-----------------------------------------------------------------------%
Please note that the equation of motion for $\phi$ 
in the Einstein frame (\ref{eomEFomega01}) is 
the same as that in the Jordan one (\ref{eom_phi01}), 
and the relation between $\ti{t}^{(i)}_{\m\n}$ and
$t^{(i)}_{\m\n}$ only differs a factor, i.e.
\begin{align}
\ti{t}^{(i)}_{\m\n} = \O^{-2} t^{(i)}_{\m\n}\,. \label{relation1}
\end{align}
Clearly, (\ref{relation1}) can be expressed in terms of quantities in the Einstein frame 
with the same form
\begin{align}
& \frac{a^2 a' \phi' -a^2 a' \phi' 
  - a^3\phi''(1-a^2)}{\phi'}\ti{\rho}_{i}
  + 6a^2 a' \ti{P}_{i} \nonumber\\
&\quad +\frac{a^3[a  \phi'^{-1}\phi''^2 (1-a^2) 
       - 3 a'\phi'']}{3 a \phi'' +3a' \phi'+a^3 \phi''}
         (3\ti{P}^{\text{(in)}}_{i}-\ti{\rho}^{\text{(in)}}_{i}) = 0\,.
\end{align}
where the prime `` $\prime$ '' denotes the derivative with respect to $\ti{t}$.
For the $\o$ field, we have 
$\o = \o(F(t)) = \o(t)$. 
From (\ref{ti_ttmn01}), 
one gets
\begin{align}%%%%%%%%%%%
\ti\r_\o  &= \ti{\mft}_{00} 
           =\frac{1}{2}\o'^2 + V(\o)\,, \nonumber\\%\label{solomega01}\\ 
\ti{P}_\o &= \ti{\mft}^{1}{}_{1} 
           = \ti{g}^{11} \ti{\mft}_{11} 
           = \frac{1}{2}\o'^2 - V(\o)\,.\label{solomega02}
\end{align}

%-----------------------------------------------------------------------%
In addition,
$\ti{\l} = \ti{\rho}^{\text{(in)}}_{i}/\ti{\rho}_i$ is the proportionality between 
ordinary matter and induced matter, and 
$\ti{\o}^{\text{(in)}} = \ti{P}^{\text{(in)}}_{i}/\ti{\rho}^{\text{(in)}}_{i}$ 
is the equation of state for induced matter in the Einstein frame. 
They are related to the corresponding quantities in the Jordan frame by 
$\ti{\l}=\O^2 \l$ and $\ti{\o}^{\text{(in)}}_{i}=\o^{\text{(in)}}_{i}$.
The EoS of non-relativistic matter and radiation
are given by $\ti{w}_{\text{m}}=0$ and $\ti{w}_{\text{r}} = 1/3$, respectively.
Therefore, we can write down the modified Friedmann equations in the Einstein frame from
(\ref{eomEFomega01}) to be
\begin{align}%%%%%%%%%%%
\ti{H}^2 &= \frac{8 \pi G}{3}\big((\ti{\bar{\r}}_{\text{M}}+ \ti\r_\o)
            + \ti{\l}\ti{\bar{\r}}_{\text{M}}\big)\,, \nonumber\\%\label{FrieqEF01}\\
\ti{H}'  &= -4 \pi G\big((\ti{\bar{\r}}_{\text{M}}+ \ti\r_\o + \ti{P}_{\text{r}} +\ti{P}_\o )
            + \ti{\l}(1 + \ti{w}^{(\text{in})})\ti{\bar{\r}}_{\text{M}}\big)\,,
\label{FrieqEF02}
\end{align}
where 
$\ti{\bar{\r}}_{\text{M}} 
= \ti{\rho}_{\text{m}}+\ti\r_{\text{r}}$ and $\ti{H}'
= \ti{a}'/\ti{a}$ is the Hubble parameter in the Einstein frame.

%------------------------------------------------------------------%
%Subsec. A - First Law in Einstein Frame
%------------------------------------------------------------------%
\subsection{First Law in Einstein Frame}

Using $\ti\r_\o$ and $\ti{P}_\o$ in (\ref{solomega02}),
the third equation in (\ref{eomEFomega01}) leads to
\begin{align}%%%%%%%%%%%
\ti\r_\o' + 3\ti{H}(\ti\r_\o + \ti{P}_\o)
= \frac{1}{\a}\o'(\ti{\rho}_{\text{M}} 
  - 3\ti{P}_{\text{M}})\,, \label{contiEFomega01}
\end{align}
where $\ti{\rho}_{\text{M}}$ and $\ti{P}_{\text{M}}$ are the energy density and pressure
of ordinary matter and disformally induced matter in the Einstein frame, given by
\begin{align}
\ti{\rho}_{\text{M}} &= \ti{\bar{\r}}_{\text{M}}
+ \ti{\l}\ti{\bar{\r}}_{\text{M}}\,, \nonumber\\
\ti{P}_{\text{M}} &= (\ti{P}_{\text{r}} +\ti{P}_\o )
	+\ti{\l} \ti{w}^{(\text{in})}\ti{\bar{\r}}_{\text{M}}\,,
\end{align}
respectively.
Since the continuity equation for matter and induced matter in
the Jordan frame holds, 
the equation 
$\ti{\na}_\m(\ti{T}^{(i)}{}^\m{_\n} + \ti{t}^{(i)\m}{_\n})$ 
is no longer zero.
It follows that
\begin{align}%%%%%%%%%%%
\ti{\rho}_{\text{M}}'+3\ti{H}(\ti{\rho}_\text{M}+\ti{P}_\text{M}) 
= -\frac{1}{\a}\o'(\ti{\rho}_\text{M}-3\ti{P}_\text{M})\,. \label{contiEFmf01}
\end{align}
where $\ti{\rho}_\text{M}=\O^{-4}\rho_\text{M}$ 
and 
$\ti{P}_\text{M}=\O^{-4}P_\text{M}$.
As (\ref{contiEFomega01}) and (\ref{contiEFmf01}) have the opposite sign, 
we can combine them to form a total conserved continuity equation, 
given by
\begin{align}%%%%%%%%%%%
\ti\r_t' + 3\ti{H}(\ti\r_t + \ti{P}_t) = 0,
\end{align}
where 
$\ti\r_t =\ti{\rho}_\text{M} + \ti\r_\o$ and $\ti{P}_t 
         = \ti{P}_\text{M} + \ti{P}_\o$.
The relations in the Einstein frame are very similar to 
those in the Jordan frame in the equilibrium picture.
Hence, thermodynamics in the Einstein frame should be 
considered as an equilibrium description.

%-----------------------------------------------------------------------%
To investigate the first law of thermodynamics in the Einstein frame, 
we can follow the similar steps shown in the Jordan one.
The apparent horizon in the new frame is
\begin{align}%%%%%%%%%%%
\rt_A = \ti{H}^{-1}\,.
\end{align}
As a result, the surface area and horizon volume become 
$\ti{A} = 4\pi\rt_A^2$ and $\ti{V} = 4\pi\rt_A^3/3$, 
respectively.
Defining the Bekenstein-Hawking entropy 
$\ti{S} = \ti{A}/4G$ and Hawking temperature 
$\ti{T} = |\ti\k_s|/2\pi$,
we have
\begin{align}%%%%%%%%%%%
\ti{T}d\ti{S} 
= 4\pi\rt_A^2\Big(\big(\ti{\bar{\r}}_{\text{M}} 
  + \ti\r_\o + \ti{P}_{\text{r}} +\ti{P}_\o \big)
  + \ti{\l}\big(1 +\ti{w}^{(\text{in})}\big)\ti{\bar{\r}}_{\text{M}}\Big)
  \Big(1 + \frac{\ti r_A^2 \ti{H}'}{2}\Big) d\ti{t}\,.
\end{align}
Using the Misner-Sharp energy $\ti{E} = \rt_A/2G$ within the apparent horizon, we get
\begin{align}%%%%%%%%%%%
d\ti{E} = - 4\pi\rt_A^2\bigg[
\Big((1+\ti{H}'\ti{r}^2_A)(\ti{\bar{\r}}_{\text{M}}
  + \ti\r_\o) + \ti{P}_{\text{r}} +\ti{P}_\o\Big)
+ \ti{\l}\Big(\big(1+\ti{H}'\ti{r}^2_A 
+ \ti{w}^{(\text{in})}\big)\ti{\bar{\r}}_{\text{M}}\Big) \bigg] d \ti t\,.
\end{align}
Combining these two equations together with the introduction of the work density 
\begin{align}
\ti{W} = \frac{1}{2}\bigg(\ti{T}^{(\text{m)}ab} 
         + \ti{T}^{(\text{r)}ab} + \ti{t}^{(\text{m)}ab} 
         + \ti{t}^{(\text{r)}ab} + \ti\mft^{ab} \bigg) \ti{h}_{ab}\,,
\end{align}
where $\ti{h}_{ab}=\text{diag}(-1,\ti{a}^2)$ with $a, b=0, 1$,
the first law in the Einstein frame is given by
\begin{align}%%%%%%%%%%%
\ti{T}d\ti{S} = -d\ti{E} + \ti{W}d\ti{V}\label{1stlawEF}
\end{align}

%------------------------------------------------------------------%
%Subsec. B - Second Law in Einstein Frame
%------------------------------------------------------------------%
\subsection{Second Law in Einstein Frame}

In the Einstein frame, we can also construct the Gibbs function
\begin{align}%%%%%%%%%%%
\ti{T}d\ti{S}_t 
= d(\ti\r_t V) + \ti{P}d\ti{V} 
= \ti{V}d\ti\r_t + (\ti\r_t + \ti{P}_t)d\ti{V}\,.
\end{align}
Similarly, we have
\begin{align}%%%%%%%%%%%
\frac{d}{d\ti{t}}(\ti{S} + \ti{S}_t) 
= -\frac{48 \pi^2 \ti{H}'}{\ti{R} \ti{H}^3}\bigg(\big(\ti{\bar{\r}}_{\text{M}} 
  + \ti\r_\o + \ti{P}_{\text{r}} +\ti{P}_\o \big)
  + \ti{\l}\big(1+ \ti{w}^{(\text{in})}\big)
    \ti{\bar{\r}}_{\text{M}}\bigg)\,.\label{Eq:entropy rate in e-frame}
\end{align}
Substituting (\ref{FrieqEF02}) into the above equation, one gets
\begin{align}%%%%%%%%%%%
\frac{d}{d\ti{t}}(\ti{S} + \ti{S}_t) = \frac{12 \pi \ti{H}'^2}{G \ti{R} \ti{H}^3}\,,
\end{align}
which is obviously always positive for the accelerating expansion of the universe,
so that
\begin{align}%%%%%%%%%%%
\frac{d}{d\ti{t}}(\ti{S} + \ti{S}_t) 
= \frac{12 \pi \ti{H}'^2}{G \ti{R} \ti{H}^3} 
  \geq 0\,,\label{2ndlawEF}
\end{align}
as expected by the second law of thermodynamics.
We can see that the solution is similar to that of the equilibrium picture in the Jordan frame.
By replacing all the variables in terms of those in the Einstein frame,
they both describe the same picture but in the different frames.
In the limit of $\ti{\l} \rightarrow 0$, the results in (\ref{1stlawEF}) and (\ref{2ndlawEF}) 
can also be reduced to those in $f(R)$ gravity without the disformal transformation in the Einstein frame.

%-----------------------------------------------------------------%
%Subsec. C - Thermodynamics Relation Between Two Frame
%------------------------------------------------------------------%
\subsection{Thermodynamics Relation Between Two Frames}

As the thermodynamic properties in the two frames are derived,
we can find the relations between the frames.
First, the Hubble parameter has the form
\begin{align}%%%%%%%%%%%
H = \sqrt{F}\bigg(\ti{H} - \frac{1}{2F}\frac{dF}{d\ti{t}}\bigg)
= \exp^{\o/\a}\bigg(\ti{H} - \frac{\o'}{\a}\bigg)\,.
\end{align}
Then, the thermodynamics quantities in terms of $H$ are expressed by
%\begin{subequations}
\begin{align}%%%%%%%%%%%
dE  &= -\frac{dH}{2GH^2} \nonumber\\
WdV &= \bigg(3 + \frac{\dot{H}}{H^2}\bigg)dE,\nonumber\\
TdS & = \bigg(2 + \frac{\dot{H}}{H^2}\bigg)dE\,.
\end{align}
%\end{subequations}
%By defining a new parameter
%\begin{align}%%%%%%%%%%%
%\m = \frac{1-\a[\o'' + (\o')^2 - \ti{H}\o']/\ti{H}'}{(1-\a\o'/\ti{H}')^2},
%\end{align}
Consequently, we obtain the relations between the Einstein and Jordan frames to be 
\begin{align}%%%%%%%%%%%
dE  &= \mu \,d\ti E\,,\nonumber\\
WdV &= \frac{\mu}{3+\ti H'/\ti{H}^2}
       \bigg(3+\frac{(\o'/\a)(\ti{H} 
                     -\o'/\a)+(\ti{H}'-\o''/\a)}{(\ti H - \o'/\a)^2}\bigg)
       \ti{W}d\ti{V}\,,\nonumber\\
TdS &= \frac{\mu}{{2+\ti H'/\ti{H}^2}}
       \bigg(2+\frac{(\o'/\a)(\ti{H}
                     -\o'/\a)+(\ti{H}'-\o''/\a)}{(\ti H - \o'/\a)^2}\bigg)
                     \ti{T}d\ti{S}\,.
\end{align}
where $\mu = -{\ti{H}^2}/{\O(\ti{H}-\o'/\a)^2}d\ti{E}$

%%%%%%%%%%%%%%%%%%%%%%%%%%%%%%%%%%%%%%%%%%%%%%%%%%%%%%%%%%%%%%%%%%%%%%%%%%%%%%%%%%%%%%%%%%%%%%%%
%-----------------------------------------------------------------------%
% Sec. V - Conclusion
%-----------------------------------------------------------------------%
\section{Conclusions}

We have studied thermodynamics in $f(R)$ gravity with the 
disformal transformation of 
$\eta_{\m\n} = A(\phi, X)g_{\m\n} + B(\phi, X)\pa_\m\f\pa_\n\f$, 
where $\f$  and $X$ are the disformal field and corresponding kinetic term, while
$A$ and $B$ are functions of $\phi$ and $X$, respectively.
Under the assumption of the Minkowski matter metric,
we have given the Friedmann equations including the disformal field.
Particularly, we have shown that the induced EoS of $w^{(\text{in})}$
depends on the disformal field (\ref{Eq:Induced EoS}).

%-----------------------------------------------------------------------%
We have verified the first and second laws of thermodynamics for $f(R)$ gravity
with the disformal transformation in the FLRW universe in the both
equilibrium and non-equilibrium pictures.
In the equilibrium picture, dark energy 
obeys the continuity equation, whereas it does not  in the non-equilibrium one.
In addition, the disformal $\l$-dependent terms appearing in the first and second laws 
arise from the disformal field, which are absent in the 
standard $f(R)$ gravity theory. 
To demonstrate these contributions, we connect our model to that of
the $f(R)$ theory without the disformal relation.
We show that with the absence of the disformal relation by setting $\l \rightarrow 0$,
our equations describing the first and second laws 
of thermodynamics can be reduced to those in the ordinary $f(R)$ gravity theory.

%-----------------------------------------------------------------------%
We have also confirmed the first and second laws of thermodynamics
for $f(R)$ gravity with the disformal transformation in the Einstein frame.
By finding the relations between quantities in the Jordan
and Einstein frames, we have shown that the contributions
from the disformal field in the Einstein frame can be expressed
as the disformal $\ti{\l}$-dependent terms with $\ti{\l}=\O^{2} \l$.
Similarly, when taking the limit of $\ti{\l} \rightarrow 0$, the equations
of the first and second laws of thermodynamics go back to  the ordinary ones in $f(R)$ gravity in the Einstein frame.

%-----------------------------------------------------------------------%
%{\color{red}
We remark that in the Jordan frame, we have both non-equilibrium and equilibrium
descriptions.  As shown in Equation \eqref{eqvsnoneq}, the change of the horizon entropy $S$ in the equilibrium
picture includes the information of both $d\hat{S}$ and $d_{i}S$ in the non-equilibrium one.
Clearly, the existence of the non-equilibrium description in the Jordan
frame but not in the Einstein frame also gives us an implication
that two frames are inequivalent.
%}

%----------------------------------------------------------------------%
%{\color{red}
	In this paper, we only consider the case that the disformal
	metric couples to matter. Thus, the effect of this coupling can be interpreted
	as an additional matter (disformally induced matter). The first law of 
	thermodynamics will be modified by adding additional matter contents
	in the both Jordan and Einstein frames.
	Also, calculations show that the second law of thermodynamics depends
	on the total energy density and pressure, and hence, can also be verified
	by taking into account additional matter contents.
	%}

%------------------------------------------------------------------------%
%{\color{red}
	Finally, it is worth noting that the equation of state, defined in
	(\ref{Eq:Induced EoS}),
	is independent of the matter contents, implying the \emph{unique} phase of the induced matter.
Furthermore, if the matter metric is not the Minkowski one, it should be interesting to 
study how the thermodynamic properties will change in both frames.
%}

%%%%%%%%%%%%%%%%%%%%%%%%%%%%%%%%%%%%%%%%%%
%\authorcontributions{Methodology, C.Q. Geng and W.C. Hsu; Calculation, W.C. Hsu, J.R. Lu and L.W. Luo; Writing---original draft, C.Q. Geng, W.C. Hsu, J.R. Lu, L.W. Luo; All authors have read and approved the final manuscript.}

%%%%%%%%%%%%%%%%%%%%%%%%%%%%%%%%%%%%%%%%%%%%%%%%
%\funding{This research was funded by National Center for Theoretical Sciences, Ministry of Science and Technology grant number MoST-104-2112-M-007-003-MY3 and MoST-107-2119-M-007-013-MY3 and Academia Sinica Career Development Award Program grant number AS-CDA-105-M06.}
%Please put funding information in this section.
%%%%%%%%%%%%%%%%%%%%%%%%%%%%%%%%%%%%%%%%%%%%%%%%
%-----------------------------------------------------------------------%
% Acknowledgments
%-----------------------------------------------------------------------%
%\section*{Acknowledgments}
\begin{acknowledgments}
We would like to thank Yong Tian for useful discussions.
The work was partially supported by National Center for Theoretical Sciences,
 Ministry of Science and Technology (MoST-104-2112-M-007-003-MY3 and MoST-107-2119-M-007-013-MY3) and
 % LWL is supported by 
 Academia Sinica Career Development Award Program (AS-CDA-105-M06).
\end{acknowledgments}

%\conflictsofinterest{The authors declare no conflict of interest.} %please add this section.
%-----------------------------------------------------------------------%
% Bibliography
%-----------------------------------------------------------------------%
%\reftitle{References}
%%\externalbibliography{yes}
%%\bibliography{ref}

\end{document}